\documentclass[conference]{IEEEtran}
\usepackage{amsfonts}
\usepackage{graphicx}
\usepackage{algorithm}
\usepackage{algorithmic}
\usepackage{amssymb}
\usepackage{amsmath}
\usepackage{setspace}
\usepackage{xcolor}
\usepackage{graphicx}
\usepackage{subfigure}
\usepackage{lipsum}
\usepackage{mathtools}
\usepackage{textcomp}

\begin{document}
\title{Adaptive Resource Management in Cognitive Radar via Deep Deterministic Policy Gradient
}

\author{\IEEEauthorblockN{Ziyang Lu,  M. Cenk Gursoy,  Chilukuri K. Mohan,  Pramod K. Varshney} 
\IEEEauthorblockA{Department of Electrical Engineering and Computer Science, Syracuse University, Syracuse NY, 13066 
\\
\{zlu112, mcgursoy, ckmohan, varshney\}@syr.edu
}}


\maketitle

\begin{abstract}
In this paper, scanning for target detection, and multi-target tracking in a cognitive radar system are considered, and adaptive radar resource management is investigated. In particular, time management for radar scanning and tracking of multiple maneuvering targets subject to budget constraints is studied with the goal to jointly maximize the tracking and scanning performances of a cognitive radar. We tackle the constrained optimization problem of allocating the dwell time to track individual targets by employing a deep deterministic policy gradient (DDPG) based reinforcement learning approach. We propose a constrained deep reinforcement learning (CDRL) algorithm that updates the DDPG neural networks and dual variables simultaneously. Numerical results show that the radar can autonomously allocate time appropriately so as to maximize the reward function without exceeding the time constraint.
\end{abstract}


\section{Introduction}
Cognitive radar represents a cutting-edge technology that integrates artificial intelligence (AI) methods to enhance its performance in complex and dynamic settings. An overview of cognitive radar systems is provided in \cite{haykin2006cognitive}, focusing on signal processing, dynamic feedback from the receiver, and preservation of the information in a cognitive radar system. Moreover, cognitive radar technology supports the development of multifunctional radar systems, which are equipped with capabilities like arbitrary waveform generation and electronic beam steering, facilitating various radar functionalities such as scanning and multi-target tracking, as detailed in \cite{charlish2020development}.

Deep reinforcement learning (DRL) has increasingly been applied to dynamic decision-making problems, demonstrating significant performance improvements as shown in \cite{mnih2015human}. Its model-free nature and capabilities to handle dynamic environments make DRL suitable for cognitive radar applications. The authors in \cite{thornton2020deep} effectively utilized DRL for spectrum allocation in multi-target tracking scenarios, significantly enhancing performance while concurrently reducing interference with existing wireless systems. Additional applications of learning-based strategies in radar systems are further explored in works \cite{selvi2020reinforcement} and \cite{meng2021deep}, illustrating the extensive applications of DRL in this field.

In this work, we formulate the problem of radar scanning and target tracking under budget constraints as a constrained Markov decision process (CMDP). Related learning-based studies on CMDP are presented in \cite{tessler2018reward}, 
\cite{pham2018optlayer}, \cite{le2019batch} and \cite{fioretto2020lagrangian}. 
However, in the literature on radar, the application of DRL to the CMDP model has not been extensively studied yet.

\subsection{Contributions}
In this work, we tackle the time allocation challenge in a multi-function cognitive radar system. The considered system performs scanning to detect newly emerging targets while simultaneously tracking previously identified/detected targets. The objective of the radar is to learn an efficient time allocation strategy that jointly maximizes its performance in both scanning and tracking. Expanding on the foundation established in our previous work \cite{10622225}, we make the following significant contributions:

\begin{itemize}
    \item We detail the scanning and tracking models in a multi-function cognitive radar system, considering the track initialization with the global nearest neighbor (GNN) algorithm and target tracking with the extended Kalman filter (EKF). A utility function is defined as a metric of the joint performance for both tracking and scanning phases, addressing the tradeoff between the two operations. 

    \item We take into account the total time budget constraint and formulate the considered problem as a constrained optimization problem.

    \item We propose a constrained deep reinforcement learning (CDRL) framework to address the constrained optimization problem. The framework involves learning the parameters of the deep deterministic policy gradient (DDPG) algorithm and the dual variable simultaneously.

    \item Performance of the proposed DDPG based CDRL framework is compared with a heuristic approach and numerical results show that the designed CDRL framework can learn an efficient time allocation strategy, complying with the predefined time budget constraint.
\end{itemize}

Major differences from \cite{10622225} include the following. In \cite{10622225}, a deep Q-network (DQN)-based DRL is used, where the action space is discretized, and dwell time decisions are made one target at a time. This is done by either operating a single DQN multiple times sequentially or using multiple DQNs in parallel. In this paper, we utilize the DDPG framework with continuous action spaces and decide on all dwell time allocations simultaneously, resulting in higher fidelity decisions and more scalable operations. Additionally, in this paper, we consider a detailed track initialization model and use the number of missed targets as the scanning performance metric. 

\section{Radar Tracking Model}
\label{sectionRadar}

This section addresses multi-target tracking in two-dimensional space with the EKF. Here, we detail the tracking model and the EKF operation to estimate the targets' states.

\subsection{Target Motion Model}

Each target's kinematic state at time $t$ is represented by the vector $\mathbf{x}_t = [x_t, y_t, \dot{x}_t, \dot{y}_t]^T$, where $(x_t, y_t)$ denotes the position coordinates and $(\dot{x}_t, \dot{y}_t)$ indicates the velocity components in the $x$ and $y$ directions. Assuming constant velocity within the radar revisit interval, the state of the target evolves as follows:
\begin{equation}
\mathbf{x_{t+1}} = \mathbf{F_t} \mathbf{x_{t}} + \mathbf{w_t},
\label{target_motion}
\end{equation}
where $\mathbf{F_t} \in \mathbb{R}^{4\times4}$ is the transition matrix defined as
\begin{equation}
	\mathbf{F_t} = \begin{bmatrix}
		1&  0&  T_t&  0\\
		0&  1&  0&  T_t\\
		0&  0&  1&  0\\
		0&  0&  0&1  \\
	\end{bmatrix}
\end{equation}
and $T_t$ refers to the revisit interval at time $t$, which is determined by the radar system for tracking a specific target \cite{schope2021constrained}. In Eq. (\ref{target_motion}), $\mathbf{w_t}$ denotes the maneuverability noise, which is assumed to be multivariate zero-mean Gaussian noise with the covariance matrix
\begin{equation}
	\mathbf{Q_t} = \begin{bmatrix}
		T_t^4/4&  0&  T_t^3/2&  0\\
		0&  T_t^4/4&  0&  T_t^3/2\\
		T_t^3/2&  0&  T_t^2&  0\\
		0&  T_t^3/2&  0&T_t^2  \\
	\end{bmatrix} \sigma_{w}^2
\end{equation}
where $\sigma_{w}^2$ is the variance of the maneuverability noise of the target at time $t$ \cite{schope2021constrained}.

\subsection{Measurement Model}

To estimate target positions, the radar obtains measurements of range $r$ and azimuth angle $\theta$, subject to measurement noise. The relationship between the state vector $\mathbf{x_t}$ and measurement vector $\mathbf{z_t}$ is characterized by a nonlinear transformation function $h(\cdot)$, expressed as
\begin{equation}
	\mathbf{z_t} = h(\mathbf{x_t}) + \mathbf{v_t} = \left[\sqrt{x_t^2+y_t^2}, \quad \text{tan}^{-1}\left(\frac{y_t}{x_t}\right)\right]^T + \mathbf{v_t}
\end{equation}
where the measurement noise vector $\mathbf{v_t} = [v_{r,t}, v_{\theta,t}]^T$ consists of the range and angular noise components. These components are assumed to follow zero-mean Gaussian distributions with respective variances $\sigma^2_{r,t}$ and $\sigma^2_{\theta,t}$. For analytical convenience, the radar is assumed to be positioned at the origin.

The measurement noise is influenced by the signal-to-noise ratio $\text{SNR}_t$ of the radar echo at time slot $t$. The value of $\text{SNR}_t$ is determined by both the radar's dwell time $\tau_t$ and the target range $r_t$, following the relationship below  \cite{schope2021constrained}, \cite{koch1999adaptive}:
\begin{equation}
	\text{SNR}_t(\tau_t, r_t) = \text{SNR}_0\left(\frac{\tau_t}{\tau_0}\right)\left(\frac{r_t}{r_0}\right)^{-4}
    \label{SNR}
\end{equation}
where $\text{SNR}_0$, $\tau_0$ and $r_0$ are the reference values of the SNR, dwell time and the distance from the target to the radar. As noted  in \cite{meikle2008modern}, the measurement noise variance scales with $\text{SNR}_t$ as
\begin{equation}
	\sigma_{\bullet,t}^2 = \frac{\sigma_{\bullet,0}^2}{\text{SNR}_t(\tau_t, r_t)}
	\label{sigma}
\end{equation}
where $\bullet \in (r, \theta)$. $\sigma_{\bullet,0}^2$ denotes the reference value of the corresponding measurement noise variance. Based on  (\ref{SNR}) and (\ref{sigma}), the measurement uncertainty decreases either when more dwell time is allocated to the target or when the target moves closer to the radar.

The nonlinear measurement function $h(\cdot)$ motivates the use of an extended Kalman filter (EKF). The EKF linearizes the measurement model through an observation matrix $\mathbf{H_t} \in \mathbb{R}^{2\times4}$, defined as the Jacobian of $h(\cdot)$:

\begin{equation}
	\mathbf{H_t} = \frac{\partial h(\cdot)}{\partial \mathbf{x}}\vert_{\mathbf{x_t}} =
	\begin{bmatrix}
		
		\frac{x_t}{\sqrt{x_t^2+y_t^2}} &  \frac{y_t}{\sqrt{x_t^2+y_t^2}} &  0&  0\\
		\frac{-y_t}{x_t^2+y_t^2}&  \frac{x_t}{x_t^2+y_t^2}&  0&  0\\

	\end{bmatrix}.
\end{equation}

Considering independent measurements, the covariance matrix of the measurements is given by
\begin{equation}
	\mathbf{R_t} =
	\begin{bmatrix}
		\sigma_{r,t}^2  & 0\\
		0               & \sigma^2_{\theta, t}
	\end{bmatrix}.
\end{equation}

\subsection{Extended Kalman Filter (EKF)} \label{subsec:Kalman}
Kalman filter is an algorithm frequently utilized for estimating the state of a process through a sequence of measurements over time \cite{welch1995introduction}. For scenarios involving nonlinear measurements, such as target tracking in radar systems, the EKF serves as the nonlinear version of the Kalman filter. The EKF operates in two phases, namely the prediction and update phases: 

\subsubsection{Prediction}
\begin{equation}
	\mathbf{\hat{x}_{t|t-1}} = \mathbf{F_t} \mathbf{x_{t-1|t-1}}
	\label{Kalman1}
\end{equation}
\begin{equation}
	\mathbf{\hat{P}_{t|t-1}} = \mathbf{F_t} \mathbf{P_{t-1|t-1}} \mathbf{F_t}^T + \mathbf{Q_t}
	\label{Kalman2}
\end{equation}
\subsubsection{Update}
\begin{equation}
	\mathbf{K_{t}} = \mathbf{\hat{P}_{t|t-1}}\mathbf{H_t}^T(\mathbf{H_t} \mathbf{\hat{P}_{t|t-1}} \mathbf{H_t}^T + \mathbf{R_t})^{-1}
	\label{Kalman3}
\end{equation}
\begin{equation}
	\mathbf{x_{t|t}} = \mathbf{\hat{x}_{t|t-1}} + \mathbf{K_{t}}(\mathbf{z_{t}} - h(\mathbf{\hat{x}_{t|t-1}}))
	\label{Kalman4}
\end{equation}
\begin{equation}
	\mathbf{P_{t|t}} = (\mathbf{I} - \mathbf{K_t}\mathbf{H_t}) \mathbf{\hat{P}_{t|t-1}}
	\label{Kalman5}
\end{equation}
where $\mathbf{\hat{x}_{t|t-1}}$ and $\mathbf{\hat{P}_{t|t-1}}$ represent the predicted state and its covariance matrix, while $\mathbf{x_{t|t}}$ and $\mathbf{P_{t|t}}$ denote the updated state estimate and its associated covariance matrix.

As noted above, the EKF operates in two stages. First, Equations (\ref{Kalman1}) and (\ref{Kalman2}) are used to compute the state prediction and its covariance matrix. Then, the filter calculates the optimal Kalman gain using (\ref{Kalman3}), which weighs the measurement innovation to update the state estimate via (\ref{Kalman4}). The covariance update follows from (\ref{Kalman5}).

The filter is initialized with $\mathbf{x_{t|t}} = \mathbf{0}$ and $\mathbf{P_{t|t}} = \mathbf{I}$, indicating no prior knowledge about the targets. Through the iterative application of these equations, the EKF progressively reduces the trace of $\mathbf{P_{t|t}}$, leading to increasingly refined state estimates.

\subsection{Tracking Cost Function} \label{subsec:trackingcost}

We consider a similar cost function as used in \cite{schope2021constrained}. Specifically, the tracking cost function at time $t$ is defined as
\begin{equation}
	c_t(T_t, \tau_t) = \text{trace}(\mathbf{E} \mathbf{P_{t|t}} \mathbf{E}^T)
	\label{cost}
\end{equation}
where
\begin{equation}
	\mathbf{E} =
	\begin{bmatrix}
		1&0&0&0\\
		0&1&0&0
	\end{bmatrix}.
\end{equation}

According to (\ref{cost}), the tracking cost function incorporates both the revisit interval \(T_t\) and the dwell time \(\tau_t\). This cost can be interpreted as the sum of the posterior estimate variances for the target's estimated positions along both the \(x\)-axis and \(y\)-axis.


\section{Radar Scanning Model}

\subsection{Uniform Circular Array (UCA) Radar}
We consider a uniform circular array (UCA) radar system. It is assumed that the scanning beams are swept around the entire circular array in a phased-array manner, providing 360\textdegree \, coverage of the scanning area. Note that radar scanning is performed in order to detect targets that may have emerged recently and not have been detected yet. 

A UCA radar can send scanning beams with a specific phase delay in between, which can steer the radar beam and uniformly sweep the entire circular scanning area. We define the total scanning time as
\begin{equation}
    \tau_{s} = \frac{360^\circ}{\phi}\tau_{beam}
    \label{beam_duration}
\end{equation}
where 
$\phi$ is the phase delay between adjacent radar beams, and $\tau_{beam}$ is the time duration of each beam.

\subsection{Measurement Acquisition Model During Scanning}
In this work, the tracking dwell times $\{\tau_t^n\}_{n=1}^N$ (for $N$ targets) are determined by a DDPG-based DRL agent and the remaining time within the revisit interval is allocated to scanning the environment for detecting newly emerging targets. Then the time duration of each beam can be calculated via (\ref{beam_duration}). The received SNR at the radar receiver during the scanning phase is denoted as $\text{SNR}_{\text{scan}}$ and can then be derived as
\begin{equation}
    \text{SNR}_{\text{scan}} 
    = \frac{P_t \tau_{beam} G_t G_r \lambda^2 \sigma}{(4\pi)^3r^4LkT_s} \label{eq:SNRscan}
\end{equation}
where $P_t$ is the transmit power, $G_t$ and $G_r$ are the transmit and receive antenna gains, $\lambda$ is the radar signal wavelength, $\sigma$ is the radar cross section of the target, $r$ is the radar-target distance, $L$ is a loss factor, $k$ is Boltzmann's constant, and $T_s$ is the system temperature.  

It is further assumed that the probability of false alarm is fixed as $P_f$. 
Via Shnidman's equation \cite{shnidman2002determination}, we can map the received signal-to-noise ratio $\text{SNR}_{\text{scan}}$ to the probability of detection $P_d$. With this characterization, a target present in the environment will be detected with probability $P_d$. 
As indicated by (\ref{eq:SNRscan}),  $\text{SNR}_{\text{scan}}$ in the scanning phase is influenced by the duration $\tau_{beam}$ of each beam and the distance $r$ between the target and the radar. The variance of the measurement is inversely proportional to $\text{SNR}_{\text{scan}}$, similarly as in (\ref{sigma}). Hence, the detection probability $P_d$ varies depending on $\tau_{beam}$ and $r$.

We futher note that in each time slot, the radar has a probability $P_f$ of receiving a measurement due to a false alarm. It is assumed that measurements resulting from false alarms are uniformly distributed across the area of interest.

\subsection{Track Initialization Model}

We utilize the global nearest neighbor (GNN) algorithm to initialize a track for a new target. Specifically, a newly received measurement is defined as associated with a previous one if the distance between the two measurements falls below a predefined threshold $T_{d}$. 

For the track initialization model, it is assumed that the radar is capable of tracking a maximum of $M$ targets simultaneously with $M$ designated storage slots for the measurements obtained from each target. At time slot $t$, the measurements are stored in the following manner:
\begin{itemize}
    \item Measurements associated with prior ones are stored in the same slot as their associated measurements.
    \item Measurements not associated with any previous data are allocated a new slot, indicating the detection of a potential new target.
    \item Slots that do not receive an associated measurement are cleared.
    \item If a slot accumulates $K$ measurements/detections, the radar will assume the occurrence of a potential new target and initialize a track for it. In this work, we set $K = 3$. Hence, a target needs to be detected consecutively three times before it starts being tracked.
\end{itemize}

The metric employed to evaluate the scanning performance is defined as the difference between the total number of targets present in the environment and the number of targets currently being tracked. Hence, the goal is to keep this difference (i.e., the number of undetected targets) small and preferably equal to zero to avoid any missed targets. 

\section{Problem Formulation}

\subsection{Objective Function}

In the previous sections, the model of radar tracking and scanning and their corresponding performance metrics have been presented. Specifically, the tracking cost of target $n$ at time $t$ is $c_t^n (\tau_t^n)$ as given in (\ref{cost}), where $\tau_t^n$ denotes the dwell time allocated for tracking target $n$ in time slot $t$. In the tracking task, the goal is to minimize the total tracking cost for all targets. On the other hand, in the scanning task, the goal is to minimize the number of existing targets that are not tracked (due to not having been detected yet).

We propose the following utility function as the indicator of the joint performance of the radar system:

\begin{equation}
    U_t(\{\tau_t^n\}_{n=1}^N) = -\sum_{n=1}^N c_t^n (\tau_t^n) - \beta N_{miss},
    \label{utility}
\end{equation}
where $N$ is the number of targets the radar system is tracking, $N_{miss}$ denotes the number of targets that are not being tracked, and $\beta$ is the tradeoff coefficient to balance the performance of tracking and scanning, which can vary depending on the specific scenarios.

We consider a time-slotted system, where the duration of each time slot is $T_0$, which is the radar revisit interval that stays fixed across different time slots. Within each time slot, the strategic allocation of the available time $T_0$ between tracking and scanning tasks and also allocation among multiple tracked targets is critical. For instance, allocating more time for tracking can reduce the tracking cost for currently tracked targets, whereas dedicating more time to scanning can reduce the number of targets that are yet to be detected. Additionally, effective allocation of time among the targets already being tracked is also crucial for minimizing the total tracking cost.  Therefore, our goal is to find an efficient time allocation policy to balance the performance of the two tasks and optimize the overall performance of the radar system.

\subsection{Constrained Optimization Problem}

Following \cite{tessler2018reward} and \cite{altman1999constrained}, we can formulate the time allocation problem in the radar system as the following constrained optimization problem:
\begin{equation}
	\begin{aligned}
		\max_{\pi}  \quad & \sum_{m=0}^\infty \left[\gamma^m U_{t+m}(\{\tau_{t+m}^n\}_{n=1}^N)\right]\\
		\textrm{s.t.} \quad & \sum_{m=0}^\infty\gamma^m\left(\sum_{n=1}^N \frac{\tau_{t+m}^n}{T_0} - \Theta_{\max}\right) \leq 0
	\end{aligned}
	\label{formulated_constrained_opt}
\end{equation}
where $U$ is the utility function defined in (\ref{utility}) and $\Theta_{\max} \in (0,1]$ is the time budget limit. Note that the time budget allocated to tracking target $n$ is defined as the ratio of dwell time $\tau^n$ of target $n$ to radar revisit interval $T_0$, i.e., $\frac{\tau^n}{T_0}$. 

As indicated in the optimization problem in (\ref{formulated_constrained_opt}), the goal is to find a time budget allocation policy $\pi$ that maximizes the discounted sum utility function while satisfying the discounted sum budget constraint. In this work, we employ deep reinforcement learning (DRL) to determine such a policy. More specifically, policy $\pi$ is determined by a deep deterministic policy gradient (DDPG) agent. Details are provided in the next section.

Before we design the DRL agent, we note that by introducing a non-negative dual variable $\lambda$ (i.e., by utilizing Lagrangian relaxation), the problem in (\ref{formulated_constrained_opt}) can be relaxed to an unconstrained optimization problem as follows: 

\begin{align}
\tiny	
 \hspace{-.25cm}\min_{\lambda_t \geq 0} \max_{\pi} \sum_{m=0}^\infty \gamma^m \Bigg[U_{t+m}(\{\tau_{t+m}^n\}_{n=1}^N)\!-\!\lambda_t\!\left(\sum_{n=1}^N \frac{\tau_{t+m}^n}{T_0} \!-\! \Theta_{\max}\!\right)\Bigg].
	\label{dual}
\end{align}

\section{Constrained Deep Reinforcement Learning for Scanning and Multi-target Tracking}

\subsection{Deep Deterministic Policy Gradient (DDPG)}
As noted above, we employ a DRL agent to allocate time for all the targets simultaneously, which indicates that the action space of the agent is high-dimensional and continuous. To address this challenge, the DDPG algorithm is utilized to handle the continuous action space.

DDPG is an algorithm in the domain of reinforcement learning (RL) that specializes in addressing continuous action spaces \cite{lillicrap2015continuous}. It is an actor-critic and model-free algorithm and combines the advantages of value-based and policy-based DRL algorithms, enabling the efficient handling of complex decision-making tasks.

There are two main components of the DDPG algorithm: the actor and the critic. The actor network performs the mapping of the states to actions, i.e., decides the optimal action to take given the current state. The critic network evaluates the actions taken by estimating the Q-values of the actions. DDPG also employs experience replay and target networks to stabilize the learning process.

\subsection{Proposed DDPG Based CDRL Framework}
Below, we describe the state and action spaces as well as the reward function of the proposed DDPG agent. 
\subsubsection{State}
State $\mathbf{s}_t$ consists of the current observation of the environment. In particular, we define the state as follows:
\begin{equation}
    \mathbf{s}_t = \Bigg[\{c_{t-1}^n(\tau_{t-1}^n)\}_{n=1}^N,\{\tau_{t-1}^n\}_{n=1}^N,\lambda_{t-1}\Bigg].
\end{equation}

The first $N$ entries $\{c_{t-1}^n(\tau_{t-1}^n)\}_{n=1}^N$ are the tracking costs of the $N$ targets in the previous time slot. In scenarios where the actual number of targets is fewer than $N$, the remaining entries of this component are populated with zeros. The subsequent $N$ entries $\{\tau_{t-1}^n\}_{n=1}^N$ are the dwell times selected again in  the previous time slot $t-1$. The last entry is the value of the dual variable in the previous time slot. Overall, the state space has a size of $2N + 1$.

\subsubsection{Action}
In DDPG, the dwell times allocated to tracking multiple targets are determined simultaneously. The action is defined as $a_t = \{\tau_{t}^n\}_{n=1}^N$, where $\tau_{t}^n \in [0, T_0]$. Hence, the dimension of the action vector is $N$. The remaining time from the budget is allocated to radar scanning. 

\subsubsection{Reward}

To solve the unconstrained optimization problem in (\ref{dual}), the reward function $r_t$ is defined as
\begin{equation}
    r_t = \Bigg[U_{t}(\{\tau_{t}^n\}_{n=1}^N)-\lambda_t\left(\sum_{n=1}^N \frac{\tau_{t}^n}{T_0} - \Theta_{\max}\right)\Bigg]
    \label{reward}
\end{equation}
where the first term on the right-hand side is the utility function and the second term is the penalty if the budget constraint is violated.   
In the proposed CDRL framework, the DDPG parameters for maximizing the reward function and the value of the dual variable $\lambda_t$ are  learned simultaneously.

\subsection{Workflow of the CDRL Framework}
\begin{algorithm}[!ht]
\footnotesize
 \caption{CDRL Algorithm}
	\begin{spacing}{1}
	\begin{algorithmic}[1]
		\STATE{Initialize the parameters of the DDPG networks with random values.}
		\STATE{Initialize state $\mathbf{s_0}$ as zero vectors and dual variable $\lambda_t$ as $\lambda_0$.}
		\FOR{time slot $t=0,1,..., T_{max}$}
		\STATE{Select action $a_t$ based on the current state              $\mathbf{s}_t$ with the actor network in the DDPG algorithm.}
            \STATE{Compute the utility function according to (\ref{utility}) with the selected action $a_t$.}
		\STATE{Compute reward function $r_t$ according to                 (\ref{reward}).}
		
		\STATE{Store the experience ($\mathbf{s_t}$, $a_t$, $r_t$, $\mathbf{s_{t+1}}$) to the DDPG experience replay buffer.}
		\STATE{Update neural networks in DDPG with experience replay and back-propagation.}
		\STATE{$\lambda_{t+1} = \max(0,\; \lambda_t - \alpha (\sum_{n=1}^N \frac{\tau_{t}^n}{T_0} - \Theta_{max}) )$.}
		\ENDFOR
	\end{algorithmic}
	\end{spacing}
	\label{algorithm1}
\end{algorithm}
The proposed CDRL algorithm is given in Algorithm \ref{algorithm1}. We simultaneously update the DDPG parameters $\theta_t^{\pi}$ and the dual variable $\lambda_t$.

The objective of the proposed CDRL algorithm is to find a solution to the problem in (\ref{dual}). By defining the reward function as in (\ref{reward}), the DRL agent is trained to optimize the discounted reward $r_t$, i.e.
\begin{equation}
    \max_{\pi} \sum_{m=0}^\infty \gamma^m \left[U_{t+m}(\{\tau_{t+m}^n\})-\lambda_t\left(\sum_{n=1}^N \frac{\tau_{t+m}^n}{T_0} - \Theta_{\max}\right)\right].
    \label{DRLjob}
\end{equation}

Let $\mathcal{L}$ represent the objective function in (\ref{DRLjob}). The dual variable $\lambda_t$ is optimized by solving the minimization problem:
\begin{equation}
    \begin{aligned}
	\lambda_{t+1} &= \max(0,\; \lambda_t - \alpha \bigtriangledown_{\lambda_t} \mathcal{L}) \\
	&= \max\left(0,\; \lambda_t + \alpha \sum_{m=0}^\infty \gamma^m \left(\sum_{n=1}^N \frac{\tau_{t+m}^n}{T_0} - \Theta_{\max}\right)\right)
	\end{aligned}
\end{equation}
where $\alpha$ is the learning rate of the dual variable. The gradient $\bigtriangledown_{\lambda_t} \mathcal{L}$ can be estimated with an additional neural network but we simplify instead as
\begin{equation}
	\lambda_{t+1} = \max\left(0,\; \lambda_t + \alpha \left(\sum_{n=1}^N \frac{\tau_{t}^n}{T_0} - \Theta_{\max}\right) \right).
\end{equation}


\section{Numerical Results and Analysis}

\subsection{Experimental Setup}
We provide the values of the simulation parameters in Table \ref{table2} below.
\begin{table}[!ht]

 \renewcommand{\arraystretch}{1}
	
	\caption{Simulation Parameters}
	\label{table1}
	\centering
	\small
	\begin{tabular}{|c||c|}
		\hline $\sigma_{r,0}^2$ ($m^2$)&16\\
		\hline $\sigma_{\theta,0}^2$ $(\text{rad}^2)$&1e-6\\
		\hline $\sigma_w$ ($(m/s^2)^2$) & 16\\
		\hline Reference distance $r_0$ ($m$) & 3000\\
		\hline Reference dwell time $\tau_0$ (s) & 1\\
		\hline Revisit interval $T_0$ (s) & 2.5\\
        \hline Tradeoff coefficient $\beta$ & 2e4 \\ 
        \hline Time budget ($\Theta_{max}$) & 0.9\\
		\hline DRL discount factor $\gamma$ & 0.9\\
		\hline DRL mini-batch size & 128\\
        \hline Replay memory buffer size & 1e6\\
		\hline Initial dual variable ($\lambda_0$) & 5000\\
		\hline Step size of dual variable ($\alpha$) & 5000\\
        
		\hline
		
	\end{tabular}
	\label{table2}
\end{table}
\subsubsection{Target Spawning Model}
In the experiment, the radar system's capacity is limited to tracking a maximum of $N = 5$ targets. To increase the diversity of the environment and validate the effectiveness of the proposed framework across various environments, we have the following general target spawning model:

\begin{itemize}
    \item Every 100 iterations, a new target can join the environment with a probability of 0.05. The initial position and velocity of the targets are generated randomly.

    \item To further diversify the operational environment, any target present for more than 3000 time slots is removed from the environment.

    \item It is assumed that the radar is monitoring an area with radius 20km.  Targets that move beyond this range are considered to have exited the environment and are not tracked by the radar.
\end{itemize}

\subsubsection{DDPG Hyperparameters}

There are four networks in the DDPG algorithm. The actor network and its target network share identical structures, each consisting of two layers with 256 and 128 neurons, respectively. ReLU is used as the activation function between the layers.

Similarly, the critic network and its target network share the same structure. Each of them has two layers, both comprising 100 neurons. ReLU is used as the activation function. The learning rates for both actor and critic networks are set to be 0.001.

\subsection{Numerical Results}

\begin{figure}[h!]
    \centering
    \includegraphics[width=.4\textwidth]{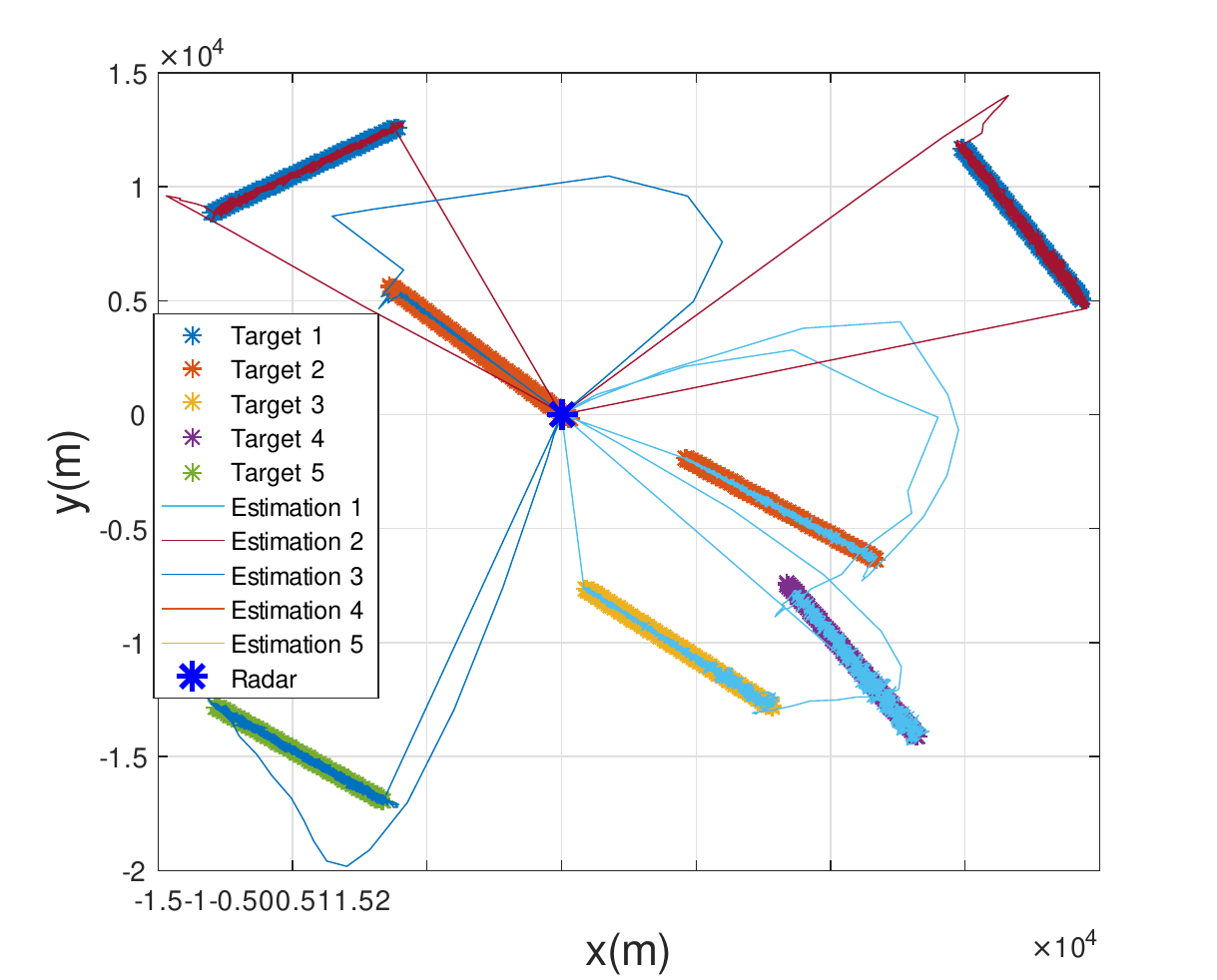}
    \vspace{-.3cm}
    \caption{Trajectories of the Targets}
    \label{traj}
\end{figure}

\begin{figure}[h!]
    \centering
    \includegraphics[width=.4\textwidth]{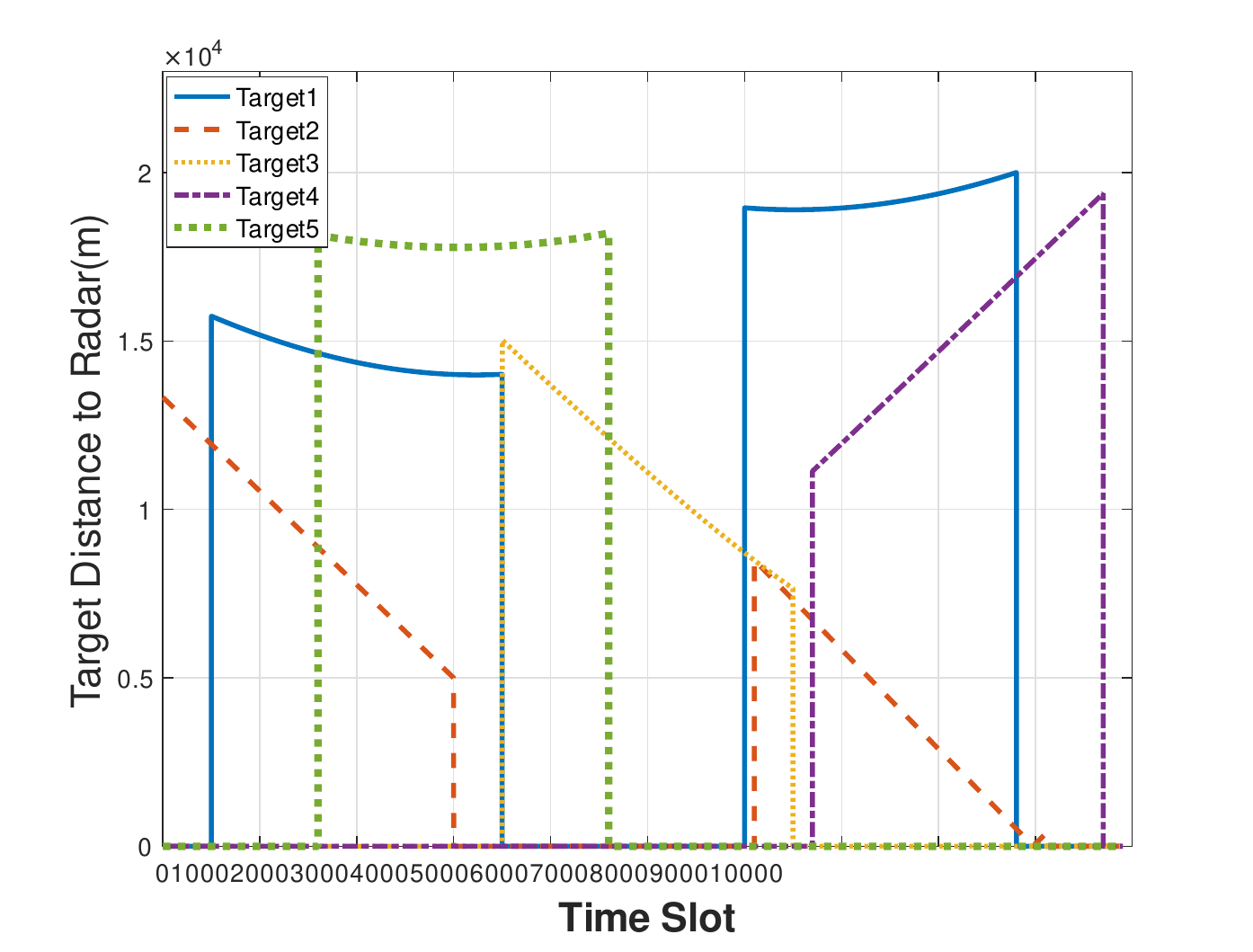}
    \vspace{-.3cm}
    \caption{Distances of the Targets to the Radar}
    \label{dist}
\end{figure}

\begin{figure}[h!]
    \centering
    \includegraphics[width=.4\textwidth]{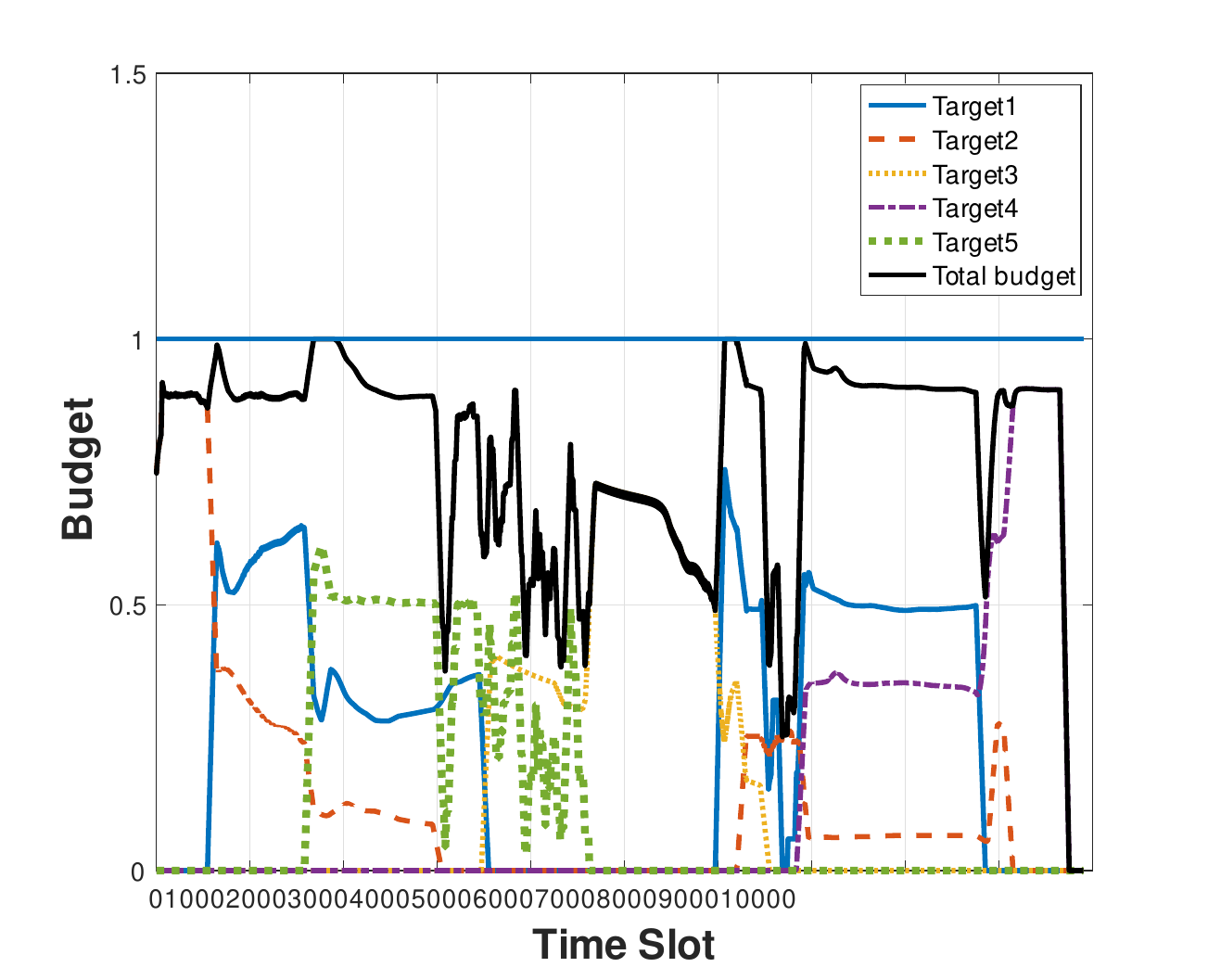}
    \vspace{-.3cm}
    \caption{Time Allocation Strategy Learned by CDRL Algorithm}
    \label{budget}
\end{figure}

Fig. \ref{traj} depicts the true locations and the estimated locations of the targets in the test environment. Radar is at the origin. During the test, both the trajectories of the targets and the time instants they are spawned in the environment are generated randomly. How the distances of the targets to the radar vary over time is shown in Fig. \ref{dist}. In Fig. \ref{traj}, we notice that the radar obtains accurate estimates of the target positions.  Specifically, we observe that the position estimates overlap with the true positions (apart from the very first initialization at the origin).

The time allocation policy learned by the proposed CDRL framework is shown in Fig. \ref{budget}, where the normalized dwell times $\{\frac{\tau_{t}^n}{T_0}\}_{n=1}^N$ allocated to tracking different targets are plotted across time slots.  The total time budget utilized for tracking is indicated as the solid black line.  When the total tracking budget is below the maximum budget of 1 (i.e., $\sum_n \tau_{t}^n < T_0$),  the remaining time budget is allocated to radar scanning. 

By comparing the target distances in Fig. \ref{dist} and the corresponding time allocation strategy in Fig. \ref{budget}, the policy learned by the CDRL framework can be summarized as follows:

\begin{itemize}
    \item  When facing scenarios with less demanding tracking tasks (e.g., fewer targets or shorter target distances), CDRL prioritizes allocating more time to the scanning task. This strategic adjustment aims to enhance the probability of capturing yet-to-be-detected targets. For instance, this behavior can be observed during time slots ranging from 5000 to 6000.

    \item Conversely, when facing scenarios with challenging tracking tasks, CDRL adjusts its strategy by dedicating more time to the tracking task, with a particular focus on allocating more dwell time for tracking distant targets. This adaptive behavior can be observed during time slots ranging from 2000 to 3000 and from 7000 to 9000.

    \item In the majority of the time slots, the total budget is constrained either at or below the predefined threshold.
\end{itemize}

\begin{figure}[h!]
    \centering
    \includegraphics[width=.4\textwidth]{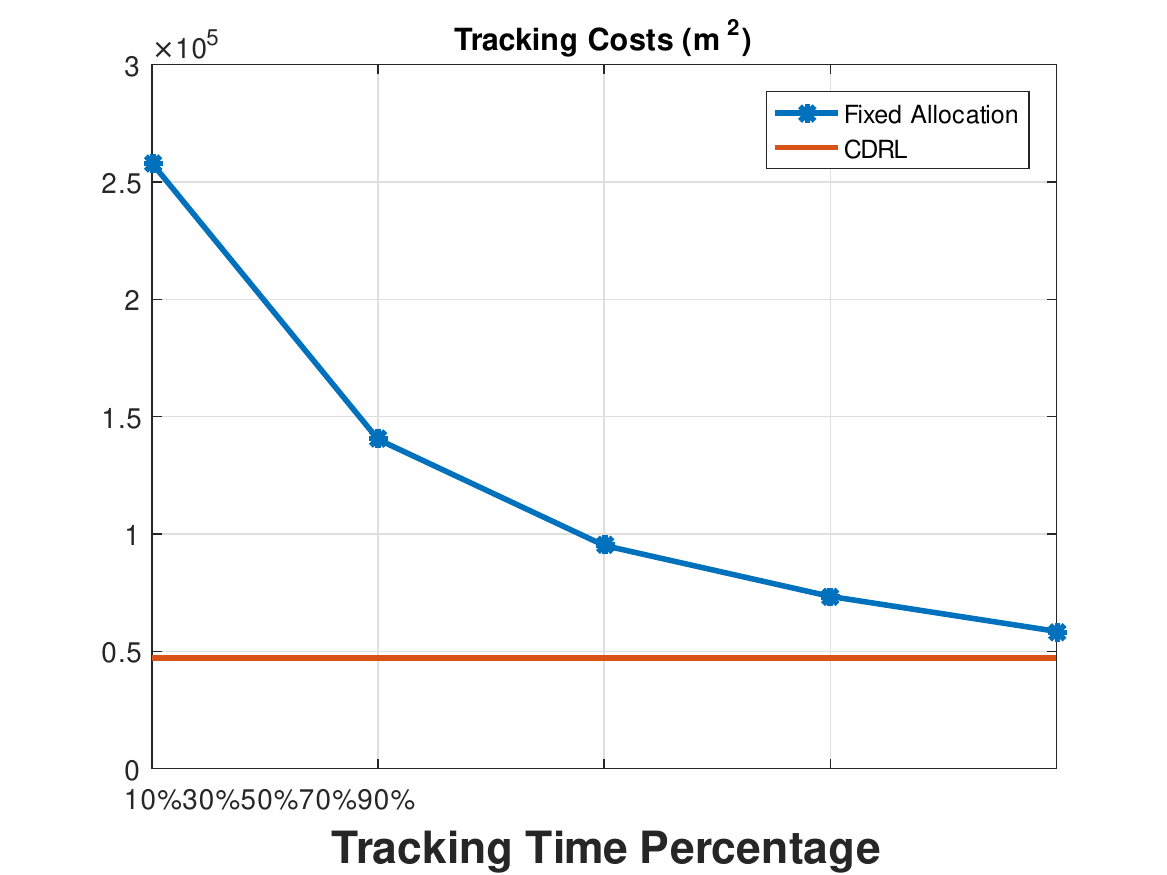}
    \vspace{-.3cm}
    \caption{Comparison of Tracking Performances}
    \label{fixtrac}
\end{figure}

\begin{figure}[h!]
    \centering
    \includegraphics[width=.4\textwidth]{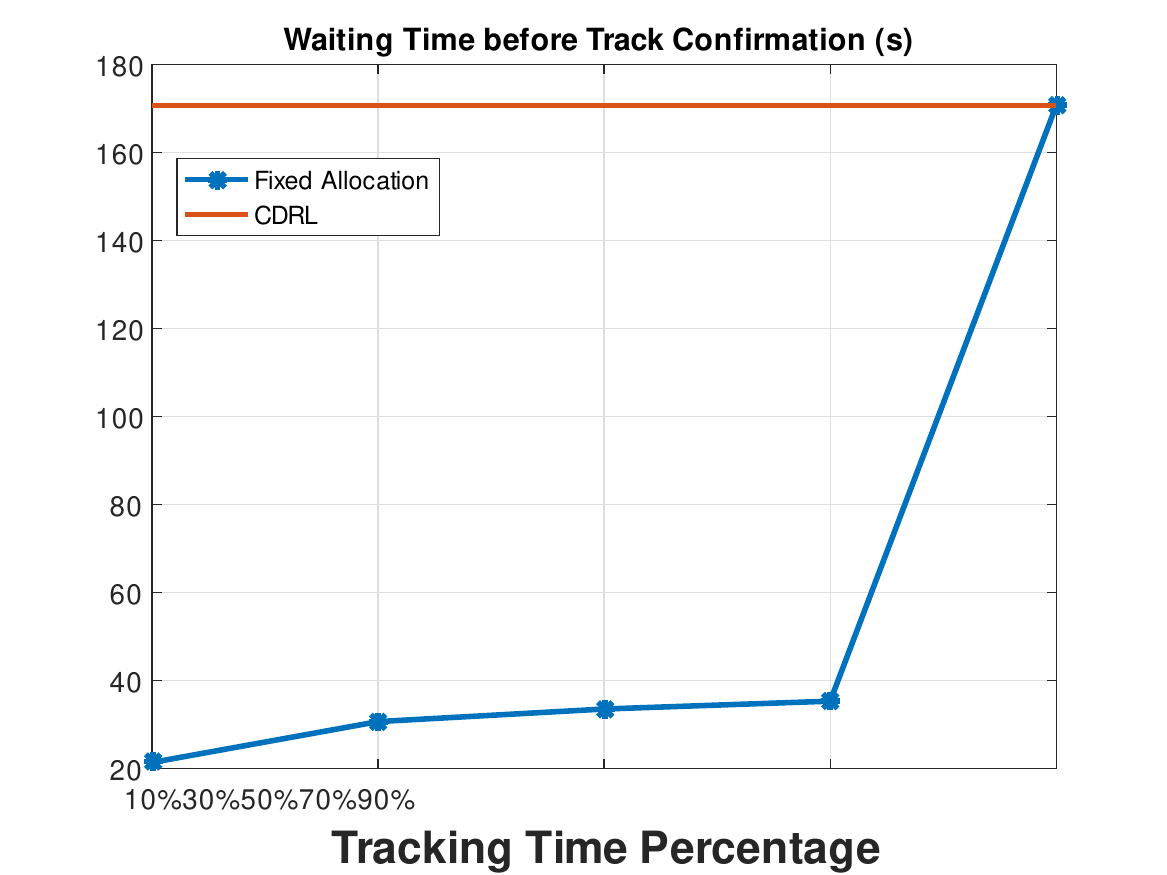}
    \vspace{-.3cm}
    \caption{Comparison of Scanning Performances}
    \label{fixscan}
\end{figure}

\begin{figure}[h!]
    \centering
    \includegraphics[width=.4\textwidth]{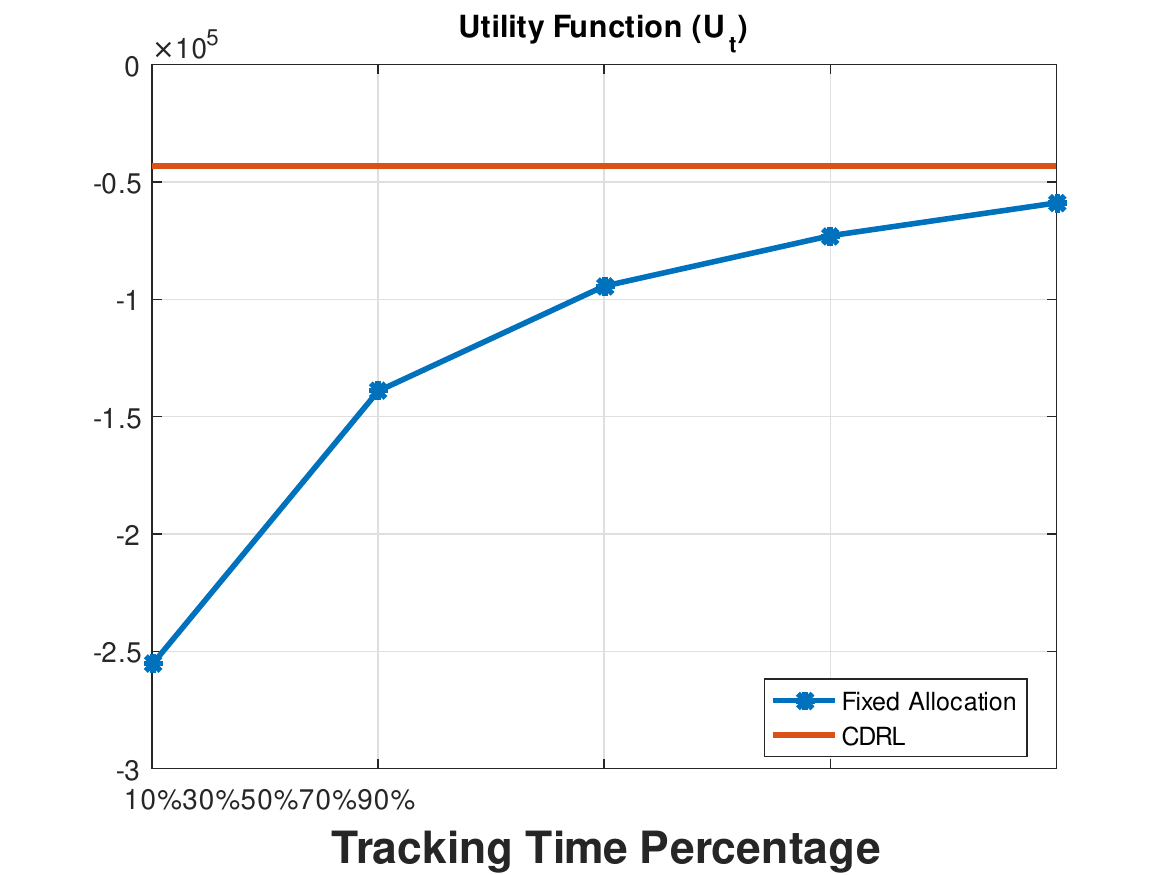}
    \vspace{-.3cm}
    \caption{Comparison of Utility Functions}
    \label{fixu}
\end{figure}

The proposed CDRL framework is further compared with a non-adaptive/fixed allocation approach. For fixed strategies, we arbitrarily select a fixed percentage of time for tracking the targets, and the dwell time is evenly distributed among all the currently tracked targets. The remaining time is allocated to scanning. The performances of tracking and scanning are depicted in Fig. \ref{fixtrac} and Fig. \ref{fixscan}, respectively. 
We observe in Fig. \ref{fixtrac} (which plots the tracking costs) that the tracking performance with fixed allocation is always worse than that of the CDRL algorithm (even when fixed $90\%$ of the time is dedicated to tracking), demonstrating the effectiveness of the dwell time allocation policy learned by the CDRL algorithm.

On the other hand, we notice in Fig. \ref{fixscan} (which plots the time it takes for track confirmation for a new target) that when the percentage of tracking time is below 90\%, it takes less time for the radar to detect a target with the fixed allocation algorithm. This is expected since fixed allocation overall provides consistently more time to scanning compared to the strategy learned by the CDRL agent. However, CDRL agent has a flexible design and its scanning performance can be improved by, for instance, increasing the value of $\beta$ in (\ref{utility}).

Fig. \ref{fixu} plots the comparison of the utility function values achieved by CDRL and fixed allocation policies. It can be seen that the cognitive radar always achieves a better performance in terms of the utility function with the proposed CDRL framework.

\section{Conclusions}
In this paper, we have proposed a DDPG based CDRL framework to address the time allocation problem for a cognitive multi-function radar. The radar system performs multiple roles: scanning the environment for undetected targets and tracking previously detected targets. The proposed framework proposes a utility function designed to effectively balance the trade-off between the scanning and tracking tasks while prioritizing targets based on their geographic locations with respect to the radar. The primary goal of the DDPG algorithm is to find efficient time allocation policies that enhance the performance of both scanning and tracking activities with a fixed total budget. Via numerical analysis, we demonstrated that the CDRL agent at the radar learns efficient strategies to optimize joint scanning and tracking performance. It maximizes the utility, intelligently allocates more time to tracking tasks with higher tracking costs and ensures sufficient time for scanning tasks, all the while adhering to budget constraints.

\bibliographystyle{IEEEtran}
\bibliography{ref}
\end{document}